\documentclass[aps,prb,amsmath]{revtex4}
 \usepackage[dvips]{graphicx}
\usepackage{amssymb}

\begin{document}

\title{Optical Identification of a DNA-Wrapped Carbon Nanotube:
Signs of Helically Broken Symmetry}
\author{Stacy E. Snyder$^\dagger$}
 \author{Slava V. Rotkin$^{\dagger\ddagger}$\thanks{Authors
 are indebted to Mr. A. A. Tsukanov for useful
discussions. This work was partially supported by DoD-ARL (grant
W911NF-06-2-0020) under Lehigh-Army Research Laboratory
Cooperative Agreement, by National Science Foundation
(CMS-0609050, NIRT), and by PA Infrastructure Technology Fund
(grant PIT-735-07).}}
 \affiliation{$^\dagger$Physics Department,
Lehigh University, 16 Memorial Dr. E., Bethlehem, PA 18015}
\affiliation{$^\ddagger$Center for Advanced Materials and
Nanotechnology, Lehigh University,  5 E. Packer Ave., Bethlehem,
PA 18015, Fax: (+1) 610-758-5730, E-mail:  rotkin@lehigh.edu}


\maketitle

High intrinsic mobility\cite{rogers} and small, biologically-compatible size make single-walled carbon nanotubes (SWNTs) in demand for the next generation of electronic devices.  Further, the wide range of available bandgaps due to changes in diameter and symmetry give SWNTs greater versatility than traditional semiconductors.\cite{charlier}  Single-stranded DNA has been employed to make these desirable properties accessible for large scale fabrication of devices.  Because single-stranded DNA can helically wrap a SWNT, forming a
stable hybrid structure \cite{zhengSci,zhengNat}, DNA/polymer
wrapping has been used to disperse bundles of intrinsically
hydrophobic SWNTs into individual tubes in aqueous solution
\cite{zhengSci,numata}.
The ability to isolate individual tubes,
make them soluble, and separate them according to symmetry
would enable fabrication of SWNT optoelectronic devices that
benefit from the unique electronic properties of specific nanotube
structures \cite{avouris}.  Envisioning optoelectronic applications of nanotubes, we investigate whether the optical properties of DNA-wrapped SWNT
materials are different than those of pristine SWNTs\cite{puller}.
 Our previous work found that bandstructures of DNA-SWNTs were indeed affected
by the charged wrap. That is, the direct optical bandgap, $E_{11}$,
decreases, but changes are fairly small
\cite{snyder}. This is consistent with the
available experimental data in standard experimental geometry in
which incident light is polarized along the SWNT axis \cite{chou}.
Here we consider optical absorption of light with perpendicular
(or circular) polarization with respect to the tube axis, which
has been measured experimentally for SWNTs dispersed using a
surfactant \cite{maruyama,kikkawa,lefebvre}.  In this geometry we
find qualitative changes in the absorption spectra of SWNTs upon
hybridization with DNA, including strong optical circular
dichroism in non-chiral SWNTs.  These optical effects are
predicted to serve as qualitative tools to directly identify the
DNA wrapping.

In general, a helical wrap may break the symmetry of a bare SWNT.
The potential of the ionized backbone of the DNA is too strong to
justify a perturbative approach. In this Communication we numerically
solve the joint Schr\"{o}dinger-Poisson equations beyond the
perturbation approximation to determine the modulation of optical
properties resulting from hybridization.

\begin{figure}[h]
\centering
\includegraphics[width=5in]{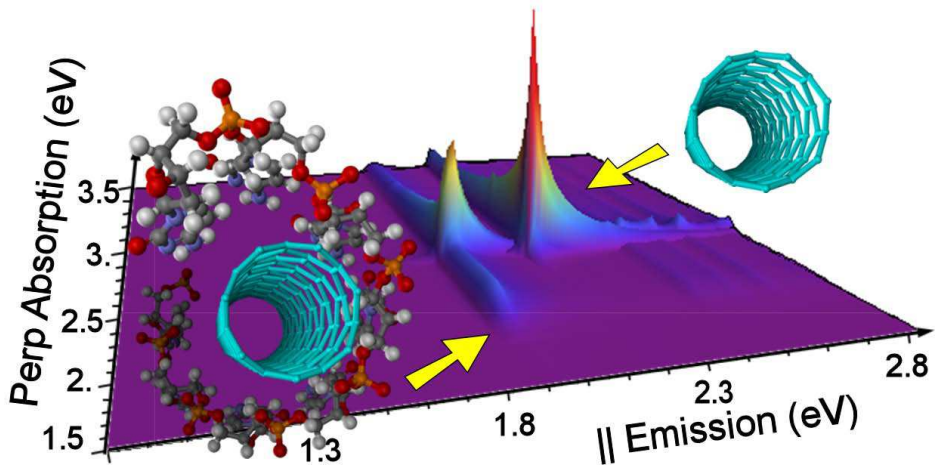}
\caption{Calculated optical strength vs. $\perp$
absorption energy and $||$ emission energy
(absorption-luminescence map) for the DNA-wrapped (7,0) nanotube (left) and the bare (7,0) nanotube (right).} \label{fig:map-1}
\end{figure}

Our simulations predict new peaks in the cross-polarized
absorption spectra of DNA-SWNTs with frequencies close to $E_{11}$
transitions, prohibited for pristine nanotubes in such polarization. In Figure
\ref{fig:map-1} we plot simulated emission along the tube axis
vs.~simulated absorption of light polarized across the tube axis
for the (7,0) SWNT with and without a DNA wrap. The figure shows a
dramatic difference in luminescence-absorption maps in the region
of the first van Hove singularity, $E_{11}$, which is explained in
the remainder of this Communication.

\begin{figure}[h]
\centering
\includegraphics[width=5in]{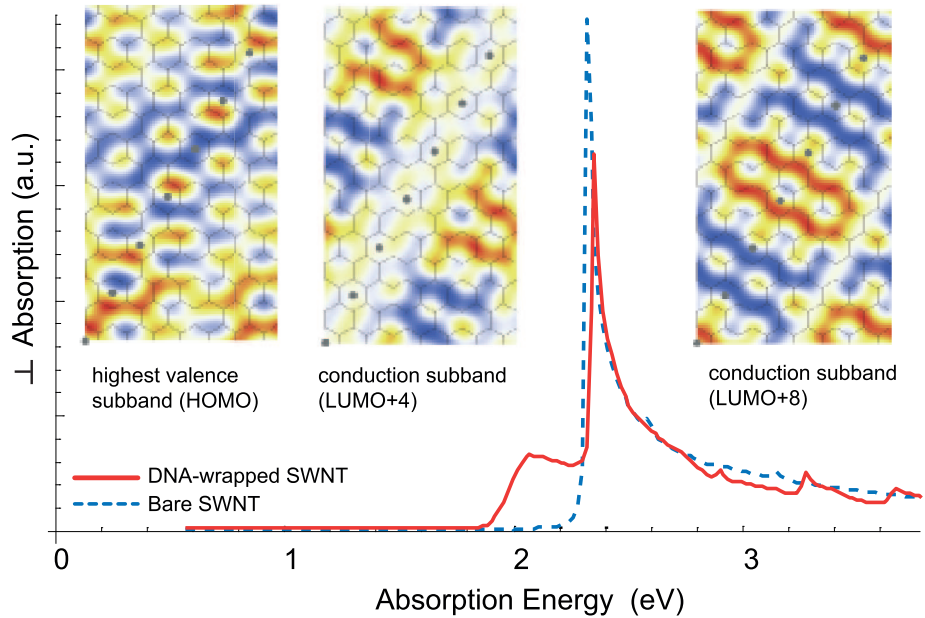}
\caption{Calculated absorption spectrum for a (7,0) SWNT with
(solid red) and without (dashed blue) a DNA wrap for perpendicular
polarization. Inset shows the potential, as a gray-scale map, and
geometry of the SWNT and DNA atoms (left) and phosphate groups
only (right) as projected onto the unravelled cylindrical tube
surface.} \label{fig:abs-compar}
\end{figure}

As we observed, absorption (and luminescence) of light with {\em
parallel} polarization should not change significantly upon DNA
hybridization \cite{snyder}. In contrast, Figure
\ref{fig:abs-compar} shows that for a semiconducting zigzag (7,0)
DNA-SWNT hybrid with the wrap geometry shown in the inset, the
optical absorption coefficients for light polarized {\em across
the SWNT
axis} 
(solid red curve) drastically
differ from the bare tube absorption in the same polarization
(dashed blue curve). The first absorption peak in
cross-polarization for the bare SWNT corresponds to $E_{12}$ and
$E_{21}$ transitions. This peak is also present for the DNA-SWNT
hybrid, although it is shifted to higher frequency. In addition, a
peak at lower frequency near that of the bare $E_{11}$ transitions
appears as a consequence of the lifting of selection rules.

For cross-polarized transitions, the selection rule for angular
momentum is $\Delta m = \pm 1$ for the bare tube. 
 This is dictated by the odd symmetry of the momentum operator and
the even symmetry of the product of wavefunctions with the same
angular momentum for the electron and the hole.

In contrast to the bare tube, the
subbands of the wrapped SWNT do not have definite angular momentum
since the electron (and hole) wave functions are helically
polarized by the Coulomb potential of the DNA. Thus, the
circularly polarized optical transitions are allowed at lower
frequency near that of the prohibited $\sim E_{11}$. The physical
interpretation of this effect is that the polarization of the
electron (hole) under the perturbation of the transverse electric
field of a nearby DNA phosphate group creates across the tube a
permanent dipole, which may be excited by the perpendicular
electric field of incident light. Overall, (negative) electron
density shifts to the opposite side of the tube and (positive)
hole density shifts towards the DNA backbone \cite{snyder}.  To
observe such an effect experimentally, one must tune to a
singularity in the optical density of states. In the bare tube
this happens only at the edges of the Brillouin zone. {\em Additional
singularities} arise with the wrapping due to subband flattening.

Within a semi-empirical orthogonal tight-binding approach, we
calculated optical absorption of a number of DNA-SWNT complexes.
This numerical approach is chosen to capture the physics of the
problem at lower computational cost.  The exciton correction is
expected to be essentially weaker for $E_{12}$ transitions due to
decay of the direct Coulomb matrix element with non-zero angular
momentum transfer $\Delta m\neq 0$ \cite{exciton}, and therefore
it is neglected in this study.  
The Hamiltonian and computational scheme are discussed in
Supplementary Materials.

The DNA backbone is modeled as a regular, infinite helix of point
charges \cite{brooks} representing the phosphate groups wrapped
around the tube (right inset in Figure \ref{fig:abs-compar}). The
angle of the helix, its position with respect the underlying
graphene lattice, the spacing between the tube and DNA charges and
its linear density are parameters of the model that can be
obtained from molecuar dynamics simulations \cite{water} or adjusted to the
experimental data. For a broad range of these
parameters, we observe similar symmetry breaking effects.

The partial absorption coefficient is calculated for a bare SWNT as
\begin{eqnarray}
\begin{array}{c}
\displaystyle \alpha_\pm(\hbar\omega,k)\propto\sum_i\sum_f
 \frac{e^2\left|\langle\psi(k, m_f, \lambda_f)|\vec p\cdot \vec e_\pm|
 \psi(k, m_i, \lambda_i)\rangle\right|^2}{m_0 \omega}
\\
\\ 
\displaystyle
 \times \frac{f(E_i(k,m_i,\lambda_i))[1-f(E_f(k,m_f,\lambda_f))]\; \Gamma}
 {[E_f(k,m_f,\lambda_f)-E_i(k,m_i,\lambda_i) -\hbar
 \omega]^2+ \Gamma^2}
 \end{array}\label{energy-single}
\end{eqnarray}
where $f(E)$ is the Fermi-Dirac function, $\Gamma$ is the inverse
lifetime, $|p_{\pm}|= |\langle\psi(k, m_i, \lambda_i)|e^{\pm i
\theta}|\psi(k, m_i\pm 1, \lambda_f)\rangle |$
 is the matrix element for a transition from an initial
state in the valence band $|k, m_i, \lambda_i\rangle$ to a final
state in the conduction band $|k, m_f, \lambda_f\rangle$ for the
case of circularly polarized incident light $\vec e_\pm$, with the
selection rule for angular momentum $\Delta m = m_f - m_i =\pm 1$.

\begin{figure}[h]
  \centering
\includegraphics[width=5in]{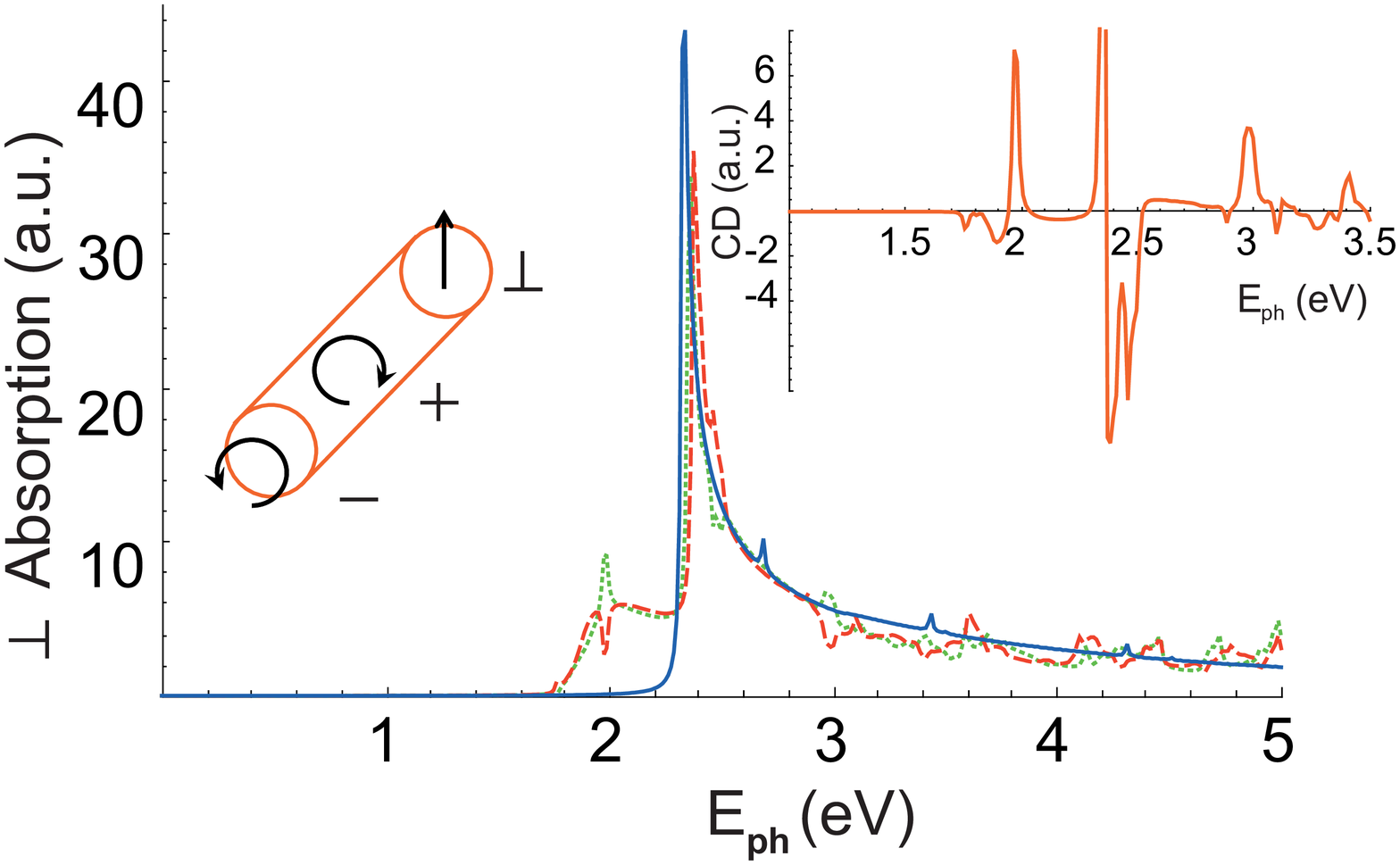}
\caption {(color online).  $\perp$ absorption of the bare (7,0)
SWNT (solid blue curve) compared to left and right polarized light
absorption in the DNA-wrapped tube (dashed red and dotted green
curves). The inset shows the circular dichroism for the hybrid.}
  \label{fig:cd}
\end{figure}

The chiral perturbation of the DNA lifts selection rules and
allows additional optical transitions in the DNA-SWNT hybrid that
were prohibited by symmetry in the bare nanotube for
cross-polarized absorption.  Such "natural" helicity of the
electron states must result in non-zero optical activity of the
material. Indeed, the optical circular dichroism spectra for a
variety of SWNT-DNA hybrids show strong dichroism for originally
achiral as well as chiral tubes (Figure \ref{fig:cd}, inset).

The total absorption spectrum is obtained by integration of the
partial absorption coefficient Eq.(\ref{energy-single}) over the
wavevector inside the first Brillouin zone $\alpha_\pm(\hbar\omega)=\int_{BZ}
\; \alpha_\pm(\hbar\omega,k)\;dk,$
 and is plotted in Figure \ref{fig:cd} for the left and right
circular polarizations $\Delta m=\pm 1$ (dashed blue and dotted
red curve, respectively).  Absorption of cross-polarized
light drastically differs for the hybrid and the bare tube (solid
green curve in Figure \ref{fig:cd}).

The difference in absorption for two polarizations gives the circular dichroism
spectrum of the DNA-SWNT hybrid: $CD= \alpha_+- \alpha_-$ (inset
of Figure \ref{fig:cd}). We stress that this circular dichroism is unrelated to the
possible chirality of the DNA. Firstly, the single-stranded DNA
does not make a clear helix on its own, without wrapping around
the SWNT. Secondly, DNA absorption is not included in this work at
all, although a recent study \cite{golovchenko} showed an
interesting DNA hypochromicity effect in the hybrids. Hence, the
optical activity must be fully attributed to the nanotube itself
and to its helical symmetry breaking. That result is further
supported by the experimental data in Ref.\cite{dukovic}.

Symmetry of the wrap is central to determine the circular
dichroism and optical response of the hybrid. We predict that
optical absorption in the perpendicular or circular polarization
may be used to detect the helical wrapping. The exact geometry of
the DNA wrap for an arbitrary tube is not yet precisely known.
Upon variation of the parameters of the wrap and/or tube
diameters, we found in our modeling that both optical effects are
almost independent of the axial or equatorial displacement of the
helical wrap along the SWNT surface without changing the helical
angle of the DNA backbone (See Suppl.). This is because the most
important factor is symmetry matching between the SWNT lattice and
the DNA backbone helical angle. For various angles and for various
tubes we predict different absorption spectra, though we believe
that a general qualitative feature of the helical symmetry
breaking, the appearance of new van Hove singularities in the
optical data, must be present.

In conclusion, we obtain optical absorption spectra and circular
dichroism for DNA-wrapped single-wall nanotubes.  In this
Communication we focus on the optical transitions for light
polarized across the nanotube axis. We predict that symmetry
lowering due to the Coulomb potential of the regular helical DNA
wrap results in qualitative changes in the cross- or circularly
polarized absorption spectrum. In particular, we predict the
appearance of a new transitions in the cross-polarized absorption
of the DNA-SWNT at frequencies substantially lower than that of
all allowed $E_{12}$ transitions in the bare tube. Therefore, with
sufficient wrapping coverage, the hybrid material is predicted to
show splitting and a shift of the lowest peak, which we suggest
can be used for experimental detection of the wrapping. A similar
effect of the symmetry breaking is predicted to result in strong
circular dichroism of the complexes.



\begin{thebibliography}{30}


\bibitem{rogers} C. Kocabas, S.-H. Hur, A. Gaur, M. Meitl, M. Shim, J. A. Rogers, \emph{Small} \textbf{2005}, \emph{1}
, 1110.

 \bibitem{charlier} J.-C. Charlier, X. Blase, S. Roche, \emph{Rev. Mod. Phys.}  \textbf{2007}, \emph{79}
, 677.

\bibitem{zhengSci} M. Zheng, A. Jagota, M. S. Strano, A. P. Santos, P. Barone, S. G. Chou, B. A. Diner, M. S. Dresselhaus, Mildred S., R. S. Mclean, G. B. Onoa, G. G. Samsonidze, E. D. Semke, M. Usrey, D. J. Walls,
\emph{Science} \textbf{2003}, \emph{302}, 1545.

\bibitem{zhengNat} M. Zheng, A. Jagota, E. D. Semke, B. A. Diner, R. S. Mclean, S. R. Lustig, R. E. Richardson, N. G. Tassi, \emph{Nat. Mater.} \textbf{2003}, \emph{2}, 338.

\bibitem{numata} M. Numata, M. Asai, K. Kaneko, A.-H. Bae, T. Hasegawa, K. Sakurai, S. Shinkai, \emph{J. Am. Chem. Soc.} \textbf{2005}, \emph{127}, 5875.

\bibitem{avouris} Ph. Avouris, \emph{Physics World} \textbf{March 2007}, 40.


\bibitem{puller} V. Puller, S. V. Rotkin, \emph{Europhys. Lett.} \textbf{2007}, \emph{77}
    , 27006.

\bibitem{snyder} S. E. Snyder, S. V. Rotkin, \emph{JETP Letters} \textbf{2006}, \emph{84}
    , 348.

\bibitem{chou} S. G. Chou, H. B. Ribeiro, E. B. Barros, A. P. Santos, D. Nezich, Ge. G. Samsonidze, C. Fantini, M. A. Pimenta, A. Jorio, F. Plentz Filho, M. S. Dresselhaus, G. Dresselhaus, R. Saito, M. Zheng, G. B. Onoa, E. D. Semke, A. K. Swan, M. S. Uenlue, B. B. Goldberg, \emph{Chem. Phys. Lett.} \textbf{2004}, \emph{397}, 296.

\bibitem{maruyama} Y. Miyauchi, M. Oba, S. Maruyama, \emph{Phys. Rev. B} \textbf{2006}, \emph{74}, 205440.

\bibitem{kikkawa} M. F. Islam, D. E. Milkie, C. L. Kane, A. G.
Yodh, J. M. Kikkawa, \emph{Phys. Rev. Lett.} \textbf{2004}, \emph{93}
, 037404.

\bibitem{lefebvre} J. Lefebvre, P. Finnie, \emph{Phys. Rev. Lett.} \textbf{2007}, \emph{98}, 167406.

\bibitem{ando} T. Ando, T. Nakanishi, R. Saito, \emph{J. Phys. Soc. Jpn.} \textbf{1998}, \emph{67}
    , 2857.

\bibitem{exciton} K. Bulashevich, S. V. Rotkin, \emph{Int. Journal of Nanoscience} \textbf{2003}, \emph{2}
    , 521.

\bibitem{brooks}B. R. Brooks, R. E. Bruccoleri, B. D. Olafson,
 D. J. States, S. Swaminathan, M. Karplus, CHARMM, \emph{J. Comp. Chem.} \textbf{1983}, \emph{24}, 187.

\bibitem{water} A. A. Tsukanov, unpublished.


\bibitem{golovchenko} M. E. Hughes, E. Brandin, J. A. Golovchenko, \emph{Nano Lett.} \textbf{2007}, \emph{7}
    , 1191.

\bibitem{dukovic} G. Dukovic, M. Balaz, P. Doak,
 N. D. Berova, M. Zheng, R. S. Mclean, L. E. Brus \emph{J. Am. Chem. Soc.} \textbf{2006}, \emph{128}, 9004.


\end{thebibliography}
\end{document}